# A microscopic derivation of nuclear collective rotation-vibration model, axially symmetric case

**P. Gulshani**

NUTECH Services, 3313 Fenwick Crescent, Mississauga, Ontario, Canada L5L 5N1
Tel. #: 647-975-8233; matlap@bell.net

## Abstract

We derive a microscopic version of the successful phenomenological hydrodynamic model of Bohr-Davydov-Faessler-Greiner for collective rotation-vibration motion of an axially symmetric deformed nucleus. The derivation is not limited to small oscillation amplitude. The nuclear Schrodinger equation is canonically transformed the to collective co-ordinates, which is then linearized using a constrained variational method. The associated constraints are imposed on the wavefunction rather than on the particle co-ordinates. The approach yields three self-consistent, time-reversal invariant, cranking-type Schrodinger equations for the rotation-vibration and intrinsic motions, and a self-consistency equation. For harmonic oscillator mean-field potentials, these equations are solved in closed forms and applied to the ground-state rotational bands in some axially-symmetric nuclei. The results are compared with the measured data.

## 1. Introduction

The phenomenological hydrodynamic rotation-vibration model of Bohr-Davydov-Faessler-Greiner (*BDFG*) [1-5] has been remarkably successful in predicting the properties of low-lying collective rotation and vibration states in deformed nuclei. The model uses adjustable parameters for the vibration and rotation masses and the first two excitation energies. Therefore, it is of interest to understand the physical nature of the assumptions, approximations, and collective flow dynamics implied by the model, and thereby have more reliable and robust model predictions as alluded to in [6]. In other words, we would like to know how the model Hamiltonian is related to the nuclear Hamiltonian.

To achieve this objective, previous studies [7-20] (refer to reference [21] for additional references) canonically transformed the nuclear Hamiltonian to a set of collective Euler angles defining the orientation of the nuclear mass quadrupole moment, a set of collective nuclear-shape co-ordinates chosen to be the diagonal components of the quadrupole moment, and a set of intrinsic co-ordinates. The nucleon co-ordinates along the quadrupole-fixed axes were subject to constraints, which partially defined the intrinsic co-ordinates. The resulting transformed Hamiltonian decomposed into a sum of three components: an intrinsic, a rotation-vibration, and an intrinsic-rotation-vibration coupled component. However, the transformation-related constraints imposed on the particle co-ordinates, the unknown complicated intrinsic co-ordinates, and the complicated nature of the collective and collective-intrinsic coupled components in the transformed Hamiltonian made it extremely difficult to perform any realistic calculation of the energy spectrum and nuclear properties. As discussed in [21], this difficulty was avoided by diagonalizing the $A$-nucleon kinetic energy plus a phenomenological potential energy in a suitably selected subspace of the harmonic oscillator irreducible representation of the non-compact group $SP(3,R)$, which is a generalization of the compact group $SU(3)$. This approach has been successful using two to six fitting parameters including a pairing interaction in one case [21].

In Section 2 of this article, we circumvent the difficulties associated with the constraints on the particle co-ordinates and the unknown intrinsic co-ordinates by transforming the nuclear Schrodinger equation (rather than the Hamiltonian) and by imposing the constraints on the wavefunction (rather than on the particle co-ordinates). We thereby deal with the space-fixed particle co-ordinates at each stage of the transformation. We choose the Euler angles such that the Coriolis coupling between the intrinsic and rotation-vibration motions vanishes. The resulting Schrodinger equation is then transformed to a set of collective shape variables. We apply a constrained variational method to linearize the transformed Schrodinger equation for an axially symmetric nucleus. In Section 3, for harmonic oscillator mean-field potentials, the derived governing equations are solved in closed forms and the solutions are simplified. In Section 4, we discuss the relationship among our, *BDFG*, and the variable-moment-of-inertia models. We also review the prescription for evaluating the predicted and measured moments of inertia. In Section 5, we compare the model predictions with the available experimental data and provide physical interpretation of the results. Section 6 concludes the article.

## 2. Derivation of microscopic rotation-vibration-intrinsic Schrodinger equation

We derive the model starting from the microscopic Bohr-Mottelson unified rotational model derived in [22] as follows.

We use the Bohr-Mottelson unified rotational-model product wavefunction:

$$\Psi_{J\mathscr{M}} = \sum_{K=-J}^{J} \mathscr{D}^{J}_{\mathscr{M}K}(\theta_s)\,\Phi_{JK}(x_{ni}) \tag{1}$$

where: $\mathscr{D}^{J}_{\mathscr{M}K}$ is the Wigner rotation matrix, $J$ , $\mathscr{M}$ , and $K$ are respectively the total angular momentum (including spin) quantum number and its $z$-components along respectively the space-fixed axes and quadrupole-moment-attached axes, whose orientations are given by the three Euler angles $\theta_s$ ( $s$ =1,2,3), and $x_{nj}$ ($n = 1,\ldots,A;\ j = 1,2,3,$ where $A =$ nuclear mass number) is a space-fixed nucleon co-ordinate. We require the non-rotational wavefunction $\Phi_{JK}$ in Eq. (1) to satisfy the constraint:

$$\hat{J}_A \,\Phi_K(x_{ni}) = 0 \tag{2}$$

which states that $\Phi_{JK}$ is a zero angular-momentum eigenstate because it is independent of $\theta_s$. However, $\Phi_{JK}$ can depend on the eigenvalues $J$ and $K$. We choose $\theta_s$ according to the criterion:

$$\frac{\partial \theta_s}{\partial x_{nj}} = \sum_{k=1}^{3} \chi^s_{jk}\, x_{nk} \tag{3}$$

(for an arbitrary, non-zero, anti-symmetric matrix $\chi^s$ for each $s$). The criterion in Eq. (3) together with the constraint in Eq. (2) makes the Coriolis interaction term (coupling the rotation and intrinsic motions in the transformed Schrodinger equation) vanish.

Performing the transformation to the co-ordinates $\theta_s$ and its conjugate angular momentum $\hat{J}_s$ on the nuclear Schrodinger equation, we obtain [22]:

$$\left(-\frac{\hbar^2}{2M}\sum_{n,j=1}^{A,3}\frac{\partial^2}{\partial x^2_{nj}} + \hat{V} - E\right)\Phi_{JK'} + \frac{1}{2}\sum_{\mathcal{M},K=-J}^{+J}\ \sum_{A,B-1}^{3} \Phi_{JK}\,\Im^{rig^{-1}}_{AB}\,\mathcal{D}^{J*}_{\mathcal{M}K'}\hat{J}_B\hat{J}_A\,\mathcal{D}^{J}_{\mathcal{M}K} = 0 \tag{4}$$

where $M$ is the nucleon mass, $\hat{V}$ is a rotationally invariant two-body interaction, $\hat{J}_A$ is the $A^{\text{th}}$ component of the total angular momentum operator along the $A^{\text{th}}$ co-ordinate axis defined by $\theta_s$, and $\Im^{rig^{-1}}_{AB}$ is the component along the axes $A$ and $B$ of the inverse of the rigid-flow kinematic moment of inertia tensor defined by:

$$\Im^{rig} \equiv Tr\left(Q\right) - Q, \qquad \text{and} \qquad Q_{jk} \equiv M\sum_{n=1}^{A} x_{nj}\,x_{nk} \tag{5}$$

where $Q$ the mass quadrupole-moment tensor. The rigid-flow kinematic moment of inertia appears in Eq. (5) because we have chosen $\theta_s$ as in Eq. (3) to eliminate the Coriolis interaction term in the transformed Schrodinger equation Eq. (4). As shown in [22], another consequence of the choice of $\theta_s$ in Eq. (3) is that the nucleon velocity field has the rigid-flow character:

$$\mathcal{V}^{rot^l}_{nj} = \left(\vec{\omega}^l \times \vec{x}_n\right)_j, \qquad \omega^l_j \equiv \Im^{rig^{-1}}_{jl} \tag{6}$$

with non-vanishing vorticity, in agreement with the rigid-flow moment of inertia in Eq. (4).

The spinless intrinsic state, rigid-flow moment of inertia and velocity field, and zero Coriolis interaction in Eqs. (2), (4), and (6) contrast with the irrotational-flow moment of inertia and two-component (rigid plus irrotational) flow velocity field in other studies [23-28], where the intrinsic state carries some angular momentum, and the resulting non-zero Coriolis interaction is determined by a shear (rotating-deforming) operator. The two models would of-course yield the same final results if the governing equations could be solved. Specifically, in our model, the interaction of the intrinsic motion with the oscillations coupled to the centrifugal stretching reduces the rigid-flow kinematic moment of inertia (as shown in Section 5), and in [23-28] studies the shear operator would be expected to increase the irrotational-flow kinematic moment of inertia.

Multiplying Eq. (4) on the left by $\mathcal{D}^J_{\mathcal{M}K}$ and integrating over $\theta_s$, we obtain [22] the effective rotation-intrinsic asymmetric rotor Schrodinger equation:

$$\begin{aligned} E\Phi_{JK} = &\left\{-\frac{\hbar^2}{2M}\sum_{nj}\frac{\partial^2}{\partial x_{nj}^2} + V + \frac{\hbar^2}{4}\left(\Im_{11}^{rig^{-1}} + \Im_{22}^{rig^{-1}}\right)\left[J(J+1)-K^2\right] + \frac{\hbar^2 K^2}{2}\Im_{33}^{rig^{-1}}\right\}\Phi_{JK} \\ &+\frac{\hbar^2}{8}\left(\Im_{11}^{rig^{-1}} - \Im_{22}^{rig^{-1}} - 2i\Im_{12}^{rig^{-1}}\right)\sqrt{(J+K+2)(J-K-1)(J+K+1)(J-K)}\,\Phi_{JK+2} \\ &+\frac{\hbar^2}{8}\left(\Im_{11}^{rig^{-1}} - \Im_{22}^{rig^{-1}} + 2i\Im_{12}^{rig^{-1}}\right)\sqrt{(J-K+2)(J+K-1)(J-K+1)(J+K)}\,\Phi_{JK-2} \\ &+\frac{\hbar^2}{4}(2K-1)\left(\Im_{13}^{rig^{-1}} - i\Im_{23}^{rig^{-1}}\right)\sqrt{(J+K+1)(J-K)}\,\Phi_{JK+1} \\ &+\frac{\hbar^2}{4}(2K+1)\left(\Im_{13}^{rig^{-1}} + i\Im_{23}^{rig^{-1}}\right)\sqrt{(J-K+1)(J+K)}\,\Phi_{JK-1} \end{aligned} \tag{7}$$

The off-diagonal elements $\Im_{AB}^{rig^{-1}}$ $(A \neq B)$ (or equivalently $Q_{AB}$) are generally small (in fact their averages generally vanish), and in this article they are ignored. In this article, we are also considering only axially symmetric deformed even-even nuclei so that $\left\langle \Im_{11}^{rig-1}\right\rangle = \left\langle \Im_{22}^{rig^{-1}}\right\rangle$ at all values of *J* and *K*, where the brackets indicate expectation values. Eq.(7) then reduces to:

$$\left\{-\frac{\hbar^2}{2M}\sum_{nj}\frac{\partial^2}{\partial x_{nj}^2} + V + \frac{\hbar^2}{2\Im_{11}^{rig}}\left[J(J+1)-K^2\right] + \frac{\hbar^2 K^2}{2\Im_{33}^{rig}}\right\}\Phi_{JK} = E\,\Phi_{JK} \tag{8}$$

We now transform Eq. (8) to the collective vibration (rigid-flow) co-ordinates defined by:

$$R_1 \equiv \frac{\Im_{11}^{rig}}{M} = \sum_{n=1}^{A}(y_n'^2 + z_n'^2)\ , \quad R_2 \equiv \frac{\Im_{22}^{rig}}{M} = \sum_{n=1}^{A}(x_n'^2 + z_n'^2)\,, \quad R_3 \equiv \frac{\Im_{33}^{rig}}{M} = \sum_{n=1}^{A}(x_n'^2 + y_n'^2) \tag{9}$$

(where the prime superscript indicates the co-ordinates along the body-fixed axes[1] given by the transformation $x_{nj} = \sum_{A=1}^{3} R_{Aj}(\theta_s)x'_{nA}$ where $R_{Aj}$ are elements of an orthogonal matrix). In transforming Eq. (8) to the co-ordinates $R_1$, $R_2$, and $R_3$ in Eq. (9), we use the product wavefunction:

[1] $R_1$, $R_2$, $R_3$ can be and are chosen independent collective co-ordinates because *A*-3 of the particle co-ordinates $x'_{nA}$ are independent of each other.

$$\Phi_{JK} = F_1(R_1)\cdot F_2(R_2)\cdot F_1(R_3)\cdot \phi_{\mathrm{JK}}(x_{nk}) \tag{10}$$

where the spherically symmetric[2] intrinsic (such as shell-model or *HFB*) wavefunction $\phi_{\mathrm{JK}}$ is subject to the constraints:

$$\frac{\partial}{\partial R_1}\phi_{\mathrm{JK}} = 0\,, \quad \frac{\partial}{\partial R_2}\phi_{\mathrm{JK}} = 0\,, \quad \frac{\partial}{\partial R_3}\phi_{\mathrm{JK}} = 0\,, \tag{11}$$

and we obtain the result [22]:

$$\begin{aligned}\sum_{n=1}^{A}\frac{\partial^2}{\partial x_{n1}^2}\Phi_{JK} =& 2A(F_1F_3\frac{dF_2}{dR_2}+F_1F_2\frac{dF_3}{dR_3})\cdot\phi_{JK}\\ &+\sum_{n=1}^{A}x_{n1}^2(4F_1F_3\frac{d^2F_2}{dR_2^2}+8F_1\frac{dF_2}{dR_2}\cdot\frac{dF_3}{dR_3}+4F_1F_2\frac{d^2F_3}{dR_3^2})\cdot\phi_{JK}\\ &+4(F_1F_3\frac{dF_2}{dR_2}+F_1F_2\frac{dF_3}{dR_3})\cdot\sum_{n=1}^{A}x_{n1}\frac{\partial}{x_{n1}}\phi_{JK}+F_1F_2F_3\cdot\sum_{n=1}^{A}\frac{\partial^2}{\partial x_{n1}^2}\phi_{JK}\end{aligned} \tag{12}$$

and similar expressions for the derivatives of $\Phi_{JK}$ with respect to $x_{n2}$ and $x_{n3}$. To derive Eq. (12) we have neglected the off-diagonal elements of $Q_{AB}$ $(A \neq B)$ and chosen the arbitrary anti-symmetric matrix $\chi^s$ in Eq. (3) to satisfy the condition:

$$\chi^s_{AB}\omega^s_{BA}C'_{BA} = -\chi^s_{AC}\omega^s_{CA}C'_{CA} \tag{13}$$

where *A*, *B*, and *C* = 1, 2 ,3 are in cyclic order, and

$$\omega^s_{AB} \equiv \sum_{j=1}^{3}\frac{\partial R_{Aj}}{\partial\theta_s}R_{Bj}\,, \qquad C'_{BC} \equiv \sum_{n=1}^{A}\left(x'^2_{nB}-x'^2_{nC}\right)$$

For $\hat{V}$ in Eq. (8) we use the harmonic-oscillator mean-field potential:

$$\hat{V} = \frac{M\omega_{\mathrm{int}}^2}{2}\sum_{n=1}^{A}r_n^2 + \frac{M\omega_{v1}^2}{2}R_1 + \frac{M\omega_{v1}^2}{2}R_2 + \frac{M\omega_{v3}^2}{2}R_3 \tag{14}$$

where the first term is the restoring potential for the intrinsic system and remaining terms are the restoring potentials for oscillations in the collective rigid-flow moment of inertia variables $R_1$,

[2] The prescriptions in Eqs. (10) and (11) extract the quadrupole-moment distribution part of the wavefunction $\Phi_{JK}$ leaving the intrinsic wavefunction $\phi_{\mathrm{JK}}$ spherically symmetric, noting that higher order deformations are found empirically to be relatively small [1,29] and hence are usually not considered at not too high angular momenta.

$R_2$, and $R_3$, and, where for application to an axially-symmetric deformed nucleus, we have chosen the oscillation frequencies $\omega_{v1}$ and $\omega_{v2}$ to be the same[3].

Substituting Eqs. (12) and (14) into Eq. (8), we obtain a transformed Schrodinger equation where the collective oscillations in each of the three spatial directions are functionally coupled to each other and to the intrinsic motion. To reduce this coupling to an algebraic coupling effectively linearizing the equation, we apply to the transformed Schrodinger equation a constrained variational method as done previously [30,31]. To achieve this goal, we take the expectation value of the equation. In the resulting equation, we then do the following: (i) we ignore, in the present model, nuclear incompressibility, (ii) we impose the following axial-symmetry conditions on the collective-oscillation wavefunctions (consistent with axially symmetric potential in Eq. (14)): $\langle F_1|R_1|F_1\rangle = \langle F_2|R_2|F_2\rangle$, $\langle F_1|\partial/\partial R_1|F_1\rangle = \langle F_2|\partial/\partial R_2|F_2\rangle$, $\langle F_1|R_1\,\partial^2/\partial^2 R_1|F_1\rangle = \langle F_2|R_2\,\partial^2/\partial^2 R_2|F_2\rangle$, i.e., the oscillations in the spatial directions 1 and 2 are assumed to be identical and in phase, at all values of *J* and *K*, (iii) we ignore any coupling between the collective oscillations in any two spatial directions, and (iv) we vary separately each of the wavefunctions $F_1^*$, $F_2^*$, $F_3^*$, and $\phi_{\mathrm{JK}}^*$ (i.e., we use the Rayleigh-Ritz variational method) subject to the normalization and energy minimization conditions: $\langle F_1|F_1\rangle = \langle F_2|F_2\rangle = \langle F_3|F_3\rangle = \langle \phi_{\mathrm{JK}}|\phi_{\mathrm{JK}}\rangle = 1$, $\partial E/\partial F_1^* = \partial E/\partial F_2^* = \partial E/\partial F_3^* = \partial E/\partial \phi_{\mathrm{JK}}^* = 0$, for arbitrary variations in $F_1^*, F_2^*, F_3^*$, and $\phi_{\mathrm{JK}}^*$. We then obtain the following three self-consistent, time-reversal invariant Schrodinger equations and a self-consistency equation:

$$\left( R_1^2 \frac{d^2}{dR_1^2} + \frac{a_+}{2} R_1 \frac{d}{dR_1} - \frac{J(J+1)-K^2}{8} + \varepsilon_1 R_1 - \frac{b_{v1}^2}{4} R_1^2 \right) |F_1\rangle = 0 \tag{15}$$

$$\left[ R_3^2 \frac{d^2}{dR_3^2} + a_- R_3 \frac{d}{dR_3} - \frac{K^2}{4} + \varepsilon_3 R_3 - \frac{b_{v3}^2}{4} R_3^2 \right] |F_3\rangle = 0 \tag{16}$$

$$\left( -\sum_{n=1}^{A} \nabla_n^2 + (\beta_1 + \beta_3)\cdot \tilde{B} + (\beta_1 - \beta_3)\cdot \tilde{B}_3 + b_{int}^2 \sum_{n=1}^{A} r_n^2 - \varepsilon_{int} \right) |\phi_{JK}\rangle = 0 \tag{17}$$

$$a_+ \cdot \beta_1 + a_- \cdot \beta_3 = \varepsilon_s \tag{18}$$

where:

[3] To satisfy the constraint in Eqs. (11), we also need to add a two-body (such as separable monopole-monopole) interactions to the right-hand-side of Eq. (14) as we did in [30,31]. In this article, we do not do so for simplicity and to gain physical insight into the results and because, using the method in [30,31], it can be shown that the effects of the constraints in Eqs. (11) are relatively small. We also satisfy partially the constraint in Eq. (2) by using a spherically-symmetric intrinsic state $\phi_{\mathrm{JK}}$ and placing a pair of nucleons of opposite spins in each of the lowest single-particle orbitals; also refer to footnote 2.

$$b_{v1} \equiv \frac{M\omega_{v1}}{\hbar}, \;\; b_{v3} \equiv \frac{M\omega_{v3}}{\hbar}, \;\; b_{int} \equiv \frac{M\omega_{int}}{\hbar}, \;\; a_1 \equiv \langle \phi_{JK} | \tilde{B} | \phi_{JK} \rangle, \;\; a_3 \equiv \langle \phi_{JK} | \tilde{B}_3 | \phi_{JK} \rangle,$$
$$a_+ \equiv a_1 + a_3, \;\; a_- \equiv a_1 - a_3, \;\; \beta_1 \equiv -4\langle F_1 | \frac{d}{dR_1} | F_1 \rangle, \;\; \beta_3 \equiv -4\langle F_3 | \frac{d}{dR_3} | F_3 \rangle \tag{19}$$

$$\tilde{B} \equiv \frac{1}{2}\sum_{n,j=1}^{A,3}\left( x_{nj}\frac{\partial}{\partial x_{nj}} + \frac{\partial}{\partial x_{nj}} x_{nj} \right), \;\; \tilde{B}_3 \equiv \frac{1}{2}\sum_{n=1}^{A}\left( x_{n3}\frac{\partial}{\partial x_{n3}} + \frac{\partial}{\partial x_{n3}} x_{n3} \right) \tag{20}$$

and $\varepsilon_1$, $\varepsilon_3$, $\varepsilon_{int}$, and $\varepsilon_s$ are functions of the reduced energy $\varepsilon \equiv 2ME/\hbar^2$ and other system parameters.

### 3. Solution of Eqs. (15), (16), (17), and (18)

The dilation-compression operator $\tilde{B}_3$ in Eq. (17) mixes states of different angular momenta. To enable us to use a single antisymmetrized intrinsic wavefuntion and, hence simplify the analysis and obtain physical insight, in this article we set $\tilde{B}_3 = \lambda \tilde{B}$ in Eq. (17) (noting that $\langle \tilde{B}_3 \rangle = a_3$, $\langle \tilde{B} \rangle = a_1$, and $\lambda = a_3/a_1$, where $\lambda < 1$ since $\tilde{B}$ includes $\tilde{B}_3$, refer to Eq. (20)). An estimate of $\lambda$ is given in Eq. (27) below in this section. We then readily obtain the solutions of Eqs. (15), (16), and (17) in closed forms[4] from the literature as discussed in [30,31], and use them to evaluate the various parameters, such as $a_1$, $\beta_1$, and $\beta_3$ in Eqs. (19). In particular, we obtain:

$$\varepsilon = 2b_c\Sigma + 4b_{v1}\left(2n_1 + a_+/2 + 2k_1\right) - \frac{2b_{v1}a_+(a_+/2-1)}{a_+/2 + 2k_1 - 1} + 2b_{v3}\left(2n_3 + a_- + 2k_3\right) - \frac{2b_{v3}a_-(a_- - 1)}{a_- + 2k_3 - 1} \tag{21}$$

where:

$$\beta_{13} \equiv (1+\lambda)\beta_1 + (1-\lambda)\beta_3, \quad b_c \equiv \sqrt{b_{int}^2 + \beta_{13}^2/4}, \quad a_3 = \lambda a_1, \quad a_+ > 2, \;\; a_- > 1 \tag{22}$$

$$2k_1 \equiv -(a_+/2-1) + \sqrt{(a_+/2-1)^2 + J(J+1)/2},$$
$$2k_3 \equiv -\left(a_- - 1\right) + \sqrt{(a_- - 1)^2 + K^2/2}, \quad \Sigma \equiv \sum_{N=0}^{N_f}\left(N + 3/2\right) \tag{23}$$

and $n_1 = 0,1,2,3,\ldots\infty$ and $n_3 = 0,1,2,3,\ldots\infty$ are the quantum numbers for the collective oscillations in respectively 1 and 3 spatial directions (and may be considered to describe the so-called beta and gamma band heads), and $\Sigma$ is the total oscillator particle-occupation number, $N$ is an oscillator-shell quantum number, $N_f$ is $N$ for the last particle-occupied (Fermi) shell.

---

[4] The solutions of Eqs. (15) and (16) given in this article differ from those of Faessler-Greiner rotation-vibration model [3,4] in three respects: (i) our solutions are not limited to small amplitude oscillations about a mean deformation, (ii) the kinematic moment of inertia is not an adjustable parameter but rather is a dynamical variable (specifically is the rigid-flow moment), and (iii) the interaction between rotation-vibration and intrinsic motions is included. Also refer to the footnote 5.

The first term on the right-hand side of Eq. (21) for the reduced total energy $\varepsilon$ is the intrinsic-system energy, the second and fourth terms are the energy eigenvalues for the collective oscillations (including the effects of rotation via the $2k_1$ and $2k_3$ parameters) in respectively 1 and 3 spatial directions, and the third and fifth terms are the energies arising from the interaction between the collective vibration displacement and intrinsic dilation-compression (i.e., from the terms $-a_+\cdot\beta_1$ and $-a_-\cdot\beta_3$ in Eqs. (17) and (19)). The third and fifth terms in Eq. (21) are observed to have always negative values. Their values increase with $J$ as the dilation-compression interaction expands the intrinsic system transferring the corresponding energy to the collective motion. This energy transfer increases the collective energy and excitation energy, reducing the rigid-flow kinematic moment of inertia. At $J=0$, third and fifth terms in Eq. (21) cancel the second and fourth terms (for $n_1=0$ and $n_3=0$), yielding the ground-state energy $\varepsilon_o = 2b_{co}\Sigma$, where $b_{co}\equiv b_c\big|_{J=0}$.

For the ground-state rotational band, for which $K=0$, we obtain:

$$\frac{b_{v1}\left[a_1(1+\lambda)/2-1\right](1+\lambda)}{a_1(1+\lambda)/2+2k_1-1}+b_{v1}(1-\lambda)\omega=\frac{a_1 b_{int}}{\sqrt{\Sigma^2-a_1^2}} \tag{24}$$

where $\omega\equiv b_{v3}/b_{v1}$. For given values of $b_{v1}$, $b_{int}$, and $\omega$ we can solve Eq. (24) together with Eq. (23) numerically. However, we obtain a closed-form solution and more physical insight by assuming that $a_1(1+\lambda)>>2$, and ignoring the relatively small term $b_{v1}(1-\lambda)\omega$ in Eq. (24), whose validity is confirmed by the results of the calculation in Section 5. Eq. (24) then yields the solution:

$$a_1=\sqrt{\frac{b_{v1}^2(1+\lambda)^4\Sigma^2-2b_{int}^2 J(J+1)}{b_{v1}^2(1+\lambda)^4+b_{int}^2(1+\lambda)^2}} \tag{25}$$

Clearly, Eq. (25) has no real solutions at and above the cut-off angular momentum $J_c$ given by:

$$J_c(J_c+1)=\frac{b_{v1}^2(1+\lambda)^4\Sigma^2}{2b_{int}^2} \quad\Rightarrow\quad J_c=-\frac{1}{2}+\frac{1}{2}\sqrt{1+\frac{2b_{v1}^2(1+\lambda)^4\Sigma^2}{b_{int}^2}} \tag{26}$$

Therefore, the rotational band is predicted to terminate at and above $J_c$ given in Eq. (26). In practice, $J$ is a discrete number and hence, at $J_c$, we can ensure that $a_+$ and $a_-$ predicted by Eqs. (19), (22), and (25) are sufficiently larger than 2 and 1 respectively as required in Eq. (22).

To obtain a physical insight into Eq. (25) and hence the results in Eq. (26), we derive the relationship $a_1 \propto \beta_{13}\Sigma \approx (1+\lambda)\beta_1\Sigma \approx 2(1+\lambda)b_{v1}\Sigma$ (since $(1-\lambda)\beta_3$ in Eq. (22) is relatively small and $\beta_1 \approx 2b_{v1}$). From the relationship, the definitions of $a_1$ and $\beta_1$ in Eq. (19), and Eq. (25), it follows that the centrifugal-stretching force (which is proportional to $J(J+1)$) increases the size of (i.e., dilates) the intrinsic system by the amount proportional to $(1+\lambda)b_{v1}\Sigma$, and reduces the mean displacement of collective system. Therefore, Eq. (25) expresses this self-consistency between the collective oscillations and motion of the intrinsic system. Thus, the centrifugal stretching $J(J+1)$ term cannot exceed the amount of intrinsic-system dilation $b_{v1}^2(1+\lambda)^4\,\Sigma^2/b_{int}^2$ required by the self-consistency between the collective displacement and intrinsic-system dilation. Eq. (34) indicates that $J_c$ is essentially proportional to $\Sigma$. Therefore, excitation of nucleons to higher angular momentum intrinsic system orbitals results in larger $\Sigma$ and hence $J_c$, as speculated in [24].

We can evaluate $\lambda$ by determining $a_3$ from its definition in Eq. (19) as we have done for $a_1$. This evaluation is straight-forward but somewhat involved in the spherical-coordinate-system representation for the intrinsic-system wavefunction $\phi_{\mathrm{JK}}$. In Cartesian co-ordinate system, we can easily solve Eq. (17) exactly without resorting to the approximation $\tilde{B}_3 = \lambda\tilde{B}$, and for $J=0$, we find:

$$\lambda = \left.\frac{a_3}{a_1}\right|_{J=0} = \left[1+(1-\omega)\cdot\frac{\Sigma_1+\Sigma_2}{2\Sigma_3}\sqrt{\frac{b_{int}^2+4b_{v1}^2}{b_{int}^2+b_{v1}^2(1-\omega)^2}}\right]^{-1} \tag{27}$$

where the $\Sigma_k \equiv \sum_{n_k=0}^{n_{fk}}(n_k+1/2)$, $(k=1,2,3)$, is the total oscillator particle-occupation number in the $k^{\text{th}}$ spatial direction, and $n_{fk}$ is the value of the oscillator quantum number $n_k$ at the Fermi level. We use Eq. (27) as an approximation for $\lambda$ at all $J$ values.

For the ground-state rotational band (i.e., for $K=0$, $2k_3=0$, $n_1=0$, and $n_3=0$), the total energy in Eq. (21) becomes:

$$\varepsilon = 2b_c\Sigma + 4b_{v1}\left(a_+/2+2k_1\right) - \frac{2b_{v1}a_+(a_+/2-1)}{a_+/2+2k_1-1} \tag{28}$$

where $b_c$, $a_+$, and $2k_1$ are given by Eqs. (19), (22), and (25). The excitation energy of a member of the ground-state rotational band is defined by:

$$\Delta E_J \equiv \frac{\hbar^2}{2M}\left[\varepsilon(J) - \varepsilon(J=0)\right] \tag{29}$$

## 4. Comparison with Faessler-Greiner model and definition of moment of inertia

Except for the terms containing the off-diagonal elements $\Im_{AB}^{rig^{-1}}$ $(A \neq B)$ and rigid-flow moment of inertia $\Im_{AA}^{rig^{-1}}$, the rotational kinetic energy term in Eq. (7) is identical to that in the Faessler-Greiner rotation-vibration model [4].

In our present model, the axial-symmetry conditions imposed, in Section 2, on the variables $R_1$ and $R_2$, their derivatives, and the products of these (i.e., the oscillations in the spatial directions 1 and 2 being identical and in-phase) are assumed to hold at all values of $J$ and $K$. That is, the shape of the nucleus remains axially symmetric as it rotates and oscillates about this shape. To describe tri-axial deviations from the axially symmetric shape, we must consider the full three tri-axial oscillation equations rather than only the two Eqs. (15) and (16). Our present model ignores nuclear incompressibility. On the other hand, the Faessler-Greiner rotation-vibration model [4] describes an incompressible, irrotational-flow[5] rotation and small-amplitude oscillations of a tri-axial nucleus about an axially-symmetric deformed equilibrium shape. That is, in the Faessler-Greiner model, $\Im_{11}^{rig-1} \neq \Im_{22}^{rig^{-1}} \neq \Im_{33}^{rig^{-1}}$, and hence the nucleus is slightly tri-axial as it rotates and vibrates.

Except for: (a) the extra terms in $a_+$ and $a_-$ in Eqs. (15) and (16) representing the coupling of the oscillations to the intrinsic motion and for the rigid-flow inertia variables, and (b) the comments made above on incompressibility, axial symmetry versus tri-axiallity, and the factor of 2 difference between Eqs. (15) and (16) arising from imposing the axial symmetry condition, Eqs. (15) and (16) resemble those in the Bohr-Davydov-Faessler-Greiner rotation-vibration model [1-5] when a transformation to the co-ordinate $R = \rho^2$ is made (refer also to footnote 4).

However, only the second and fourth terms in Eq. (21) have some resemblance to the terms in the energy eigenvalue in the Faessler-Greiner model [4]. Even these terms are different since angular momentum appears under the square-root sign (in $2k_1$ in Eq. (23)). The main reason for this radical sign is that we do not limit the analysis to small amplitude oscillations.

For large values of $a_1$ and hence $a_+$, we may expand the energies in Eqs. (28) and (29) in powers of $J(J+1)$ to obtain the well-known phenomenological Variable-Moment-of-Inertia model expression [5,33,34-37]:

[5] However, the constant factors in the rotation and vibration masses used in the Faessler-Greiner model are not those for irrotational flow but are rather fitted to the measured excitation energies of the first excited $2^+$ states. The impact of this course of action and its possible inconsistency need to be studied because the measured inertia masses do not have irrotational-flow character. The kinematic moment of inertia in the Faessler-Greiner model is proportional to the square of the deformation parameter, whereas the rigid-flow kinematic moment of inertia in our model is insensitive to the small value of the deformation.

$$\begin{aligned}\Delta E_J &\equiv \frac{\hbar^2}{2M}\left[\varepsilon(J) - \varepsilon(J=0)\right] \\ &= AJ(J+1) - BJ^2(J+1)^2 + CJ^3(J+1)^3 + .... \\ &\equiv \frac{\hbar^2 J(J+1)}{2\mathfrak{I}_J}\end{aligned} \tag{30}$$

where the coefficients *A*, *B*, *C*, etc. are functions of $b_{v1}$, $b_{int}$, $\omega$, and $\Sigma$, and the moment of inertia $\mathfrak{I}_J$ is a function of $b_{v1}$, $b_{int}$, $\omega$, $\Sigma$, and *J*. An expansion resembling that in Eq. (30) was also obtained in the Faessler-Greiner rotation-vibration model as a result of including the vibration-rotation interaction (a second-order term in their expansion in the deformation parameters) [4].

Generally, the moment of inertia $\mathfrak{I}_J$ for a given member of rotational band with angular momentum *J* is defined by [37]:

$$\begin{aligned}\Delta E_J - \Delta E_{J-2} &\equiv \frac{\hbar^2}{2\mathfrak{I}_J}\left[J(J+1) - (J-2)(J-1)\right] \\ &= \frac{\hbar^2}{2\mathfrak{I}_J}(4J-2)\end{aligned} \tag{31}$$

Therefore,

$$\frac{2\mathfrak{I}_J}{\hbar^2} = \frac{4J-2}{\Delta E_J - \Delta E_{J-2}} \quad (MeV)^{-1} \tag{32}$$

where $\Delta E_J$ is either the predicted or measured excitation energy.

## 5. Application of model to $^{8}_{4}Be$, $^{12}_{6}C$, $^{20}_{10}Ne$, $^{24}_{12}Mg$, $^{28}_{14}Si$, $^{162}_{66}Dy$, and $^{168}_{68}Er$

In this section, we apply the model developed in Section 3 to nuclei $^{8}_{4}Be$, $^{12}_{6}C$, $^{20}_{10}Ne$, $^{24}_{12}Mg$, $^{28}_{14}Si$, $^{162}_{66}Dy$, and $^{168}_{68}Er$. For each of the nuclei, we determine the ground-state deformation parameter and $\Sigma$ in Eq. (30) from its Nilsson's self-consistent deformed-oscillator particle configuration [38]. The nucleus's oblate or prolate shape is determined from the sign of the measured intrinsic quadrupole moment [39,40,41] and, for consistency, compared with that obtained in Hartree-Fock (*HF*) and group-theoretic calculations [42-51]. We have chosen the values of the collective-oscillator frequencies $b_{int}$ and $b_{v1}$ in Eq. (19) to match, as closely as possible, the predicted excitation energy (in Eq. (28)) of the first excited $2^+$ state in the ground-state rotational band to the measured excitation energy, while ensuring that the values of $a_+$ and $a_-$ predicted by Eqs. (19), (22) and (25) are sufficiently larger than 2 and 1 respectively (as required in Eq. (22)). To calculate the electric quadrupole moment and *B*(*E*2), we have determined the frequency ratio $\omega \equiv b_{v3}/b_{v1}$ and hence $b_{v3}$ in Eq. (19) from equating the electric quadrupole moment of the nuclear ground state predicted by the model to that predicted by the Nilsson's self-consistent deformed-oscillator model using the deformation parameter predicted by the Nilsson's model [38].

The results of the analysis show that the approximations used to obtain Eq. (25) from Eq. (24) are justified.

The calculation shows that the predicted excitation energy is significantly reduced by the negative term in Eq. (28), which derives from the dilation-compression interaction term $-a_+ \cdot \beta_1$ as discussed in Section 3. However, its values increases sufficiently with *J* to cause the excitation energy to increase with *J*. Therefore, this interaction plays an important role in determining the excitation energy.

The results in Tables 1 and 2 show that the model predicts reasonably well the excitation energy $\Delta E_J$ (within 20% for ${}^{8}_{4}Be$ and 16% for ${}^{20}_{10}Ne$, and within 5% or less for the other nuclei in Tables 1 and 2). For ${}^{162}_{66}Dy$ in Table 2, the predicted excitation energy is within 20% or less of the measured excitation energy. The excitation energy is progressively overpredicted with *J*. This difference may be due to the back-bending phenomenon in the experiment, which causes the moment of inertia to increase reducing the excitation energy. Table 2 shows that the measurement-inferred moment of inertia increases with $J$ more than the predicted moment of inertia. The source of this discrepancy may be due to the neglect of triaxiallity, compressibility, and other terms in the model and their impact on the collective-intrinsic interaction (recalling from Section 3 that this interaction has significant impact on the excitation energy). The model features will be explored in a future study.

For ${}^{168}_{68}Er$ in Table 2, the agreement between the predicted and measured excitation energies is excellent, and is close to that predicted by the Faessler-Greiner rotation-vibration model [56]. Not that the Faessler-Greiner model includes the rotation-vibration interaction arising from triaxiallity, whereas our model excludes this interaction. This difference in the two models may be traceable to the difference in the kinematic moments of inertia in the two models, which is discussed in the footnote 5.

The rotational-band cut-off angular momentum $J_c$ for all the nuclei is predicted at the same value above which no experimentally observed excited state belonging to the ground-state rotational band can be found.

The (dynamic) moment of inertia is generally reasonably well predicted without using any pairing interaction. The measured and predicted moments of inertia increase gradually with *J* even in the case of ${}^{162}_{66}Dy$ where the back-bending effect may be at work. They are a factor of two or less (going from the light to heavier nuclei) smaller than the rigid-flow[6] moment of inertia. As discussed in Section 3, this result indicates that the model dilation-compression interaction (which derives from the centrifugal stretching part of the rotation) between the rotation-oscillation and intrinsic motions expands the intrinsic system. The transfer of the associated energy to the collective motion increases the excitation energy and hence reduces the rigid-flow kinematic moment of inertia, and alters the rigid-flow kinematic velocity field.

A similar reduction in the moment of inertia without using a pairing interaction has also been obtained in the Faessler-Greiner model [4] and $SP(3,R)$ model [21]. On the other hand, a number of previous studies, such as that in [57], have used a pairing interaction to reduce the cranking-model-predicted rigid-flow moment of inertia to that observed experimentally.

Tables 1 and 2 show that the measured and predicted moments of inertia for most of the nuclei are four and many-times-more smaller than the rigid-body moment of inertia (going from light to heavier nuclei), as expected.

The predicted ground-state quadrupole moment $eQ_o$ is reasonably close to the measured value. For prolate nuclei, $eQ_o$ increases with $J$ from its $J = 0$ value to higher prolate values as a consequence of centrifugal stretching (i.e., $J(J+1)$ term in Eqs. (15) and (23). This stretching occurs only in the $R_1$ (i.e., $z$ axis) direction (refer to Eq. (9)) because, in the ground-state rotational band, $K$=0 and $2k_3 = 0$ and hence no centrifugal stretching in the $R_3$ (i.e., y axis) direction can occur (as seen from Eqs. (9), (16) and (23)). Therefore, in oblate nuclei ${}^{12}_{6}C$ and ${}^{28}_{14}Si$ in their ground-state rotational band, the quadrupole moment transitions, at some higher $J$ values, to a prolate shape. Table 1 presents the available measured $eQ_o$ and electric quadrupole probability transition $B(E2)$ values. In most cases, the predicted $eQ_o$ and $B(E2)$ are somewhat close to the measured values when the measurement uncertainties are considered. However, we expect discrepancies between the predicted and measured variations in $eQ_o$ and hence $B(E2)$ with $J$. In particular, the model predicts that, in all the nuclei considered, the $B(E2)$ increases monotonically with $J$ in prolate nuclei, whereas the measured $B(E2)$ in some nuclei seems to increase at low values of $J$ and decrease at higher values of $J$. Possible reasons for this discrepancy may be: (i) the neglect of nuclear incompressibility and tri-axiallity and the resulting band mixing, discussed in Section 4, (ii) the neglect of terms in the governing Eqs. (15) and (16) that couple the oscillations in a pair of spatial directions, and (iii) partial satisfaction of the zero angular momentum constraint in Eq. (2).

## 6. Concluding remarks

In this article, we have derived a microscopic version of the remarkably successful phenomenological hydrodynamic Bohr-Davydov-Faessler-Greiner nuclear rotation-vibration model. The microscopic model accounts completely for the effects of the Coriolis interaction, resulting in the rigid-flow kinematic moment of inertia, nucleon velocity field, and vibration mass. The model defines a dilation-compression operator for the interaction between rotation-vibration and intrinsic motion, and a rotational-band cut-off angular momentum.

[6] Note that rigid-flow does not required the nucleons to be frozen at various locations as would be the case in a rigid body.

The present model describes a compressible nucleus that remains strictly axially symmetric while it rotates and oscillates about this shape. On the other hand, the Faessler-Greiner rotation-vibration model describes the collective rotation and small-amplitude oscillations of an incompressible irrotational tri-axial nucleus about an axially-symmetric deformed equilibrium shape.

The model is derived by a canonical transformation of the nuclear Schrodinger equation to collective rotation angles chosen such that the Coriolis interaction term in the transformed Schrodinger equation vanishes, yielding a tri-axial rigid-flow rotor Schrodinger equation. This equation is then transformed to collective vibration co-ordinates chosen to be the three components of the rigid-flow moment of inertia. The resulting equation is then linearized using a constrained variational method to obtain, for an axially symmetric nucleus, three coupled self-consistent, time-reversal invariant cranking-type Schrodinger equations for the intrinsic and two rotation-vibration motions, and a self-consistency equation. We are at liberty to choose any (such a shell or *HFB*) isotropic intrinsic wavefunction, subject to zero-angular momentum and zero collective-vibration displacement constraints.

The above transformations differ from the those in previous studies in the following ways: (i) the Schrodinger equation rather the Hamiltonian is transformed, (ii) the associated constraints are imposed on the wavefunction rather on the nucleon co-ordinates, (iii) one deals with the space-fixed particle coordinates and avoids intractable intrinsic co-ordinates and constraints, and (iv) the Coriolis interaction term in the transformed Schrodinger equation is eliminated by a judicious choice of the rotation angles.

For harmonic oscillator-type mean-field potentials for the vibration and intrinsic systems, and for an axially symmetric nucleus, we solve the model cranking-type Schrodiner equations and self-consistency equation in closed forms. The solutions are simplified to facilitate physical understanding, and in particular to determine an analytic expression for rotational-band cut-off angular momentum. (The simplifications are justified by the results of the calculations.) The simplified equations, with the intrinsic and collective oscillator potential energy strengths as two adjustable parameters, are applied to the ground-state rotational band of ${}^{8}_{4}Be$, ${}^{12}_{6}C$, ${}^{20}_{10}Ne$, ${}^{24}_{12}Mg$, ${}^{28}_{14}Si$, ${}^{162}_{66}Dy$, and ${}^{168}_{68}Er$. The results are encouraging.

The excitation energies are reasonably well predicted. For ${}^{168}_{68}Er$ and ${}^{162}_{66}Dy$ (with the exception of the back-bending effect), the agreement between the predicted and measured excitation energies is excellent. The cut-off angular momentum is well predicted.

The measured and predicted dynamic moments of inertia agree reasonably closely without using any pairing interaction. They are a factor of two or less smaller than the rigid-flow kinematic moment of inertia. The calculation results show that dilation-compression interaction between the rotation-oscillation and intrinsic motions increases the size of the intrinsic system transferring the corresponding energy to the collective motion. This increases the excitation energy and hence reduces the rigid-flow kinematic moment of inertia and alters the rigid-flow nucleon velocity field. Therefore, the model offers a mechanism for reducing the rigid-flow moment of inertia without using a pairing interaction. In contrast, pairing has been used in cranking-model calculations to achieve this reduction. The dilation-compression mechanism for reducing the rigid-flow kinematic moment of inertia in our model also differs from the shearing mechanism used in other transformation-related models to increase the irrotational-flow kinematic moment of inertia.

The predicted and measured moments of inertia are much smaller than the rigid-body moment of inertia.

The predicted quadrupole moments and $B(E2)$'s agree reasonably well within the measurement uncertainties with the available measured data. However, for a more accurate variations of the predicted $eQ_o$ and $B(E2)$ with $J$, we need to examine the effects of (i) nuclear incompressibility, (ii) tri-axiallity, which causes band mixing, (iii) the neglected terms in the governing equations that couple the oscillations in a pair of spatial directions, and (iv) imposing more rigorously the zero angular momentum constraint.

In future endeavours, we intend to study the impact of the model features (i) to (iv) listed above. We also intend to apply the model to the ground- and excited-state rotational bands, and examine band mixing and the phenomena at high angular momentum in axial nuclei in the light, rare-earth, and actinide regions. We plan to generalize the model to the tri-axial case, and study the above phenomena.

**Table 1. Predicted/measured excitation energy ($\Delta E_J$), Cut-off *J* ($J_c$), moments of inertia ($\mathfrak{I}_J$), $eQ_o$, *B*(*E*2)**

| | ***J*** | $\Delta E_J$ ($MeV$) **model/exp** | $2\mathfrak{I}_J/\hbar^2$ $(MeV)^{-1}$ **model/exp** | $2\mathfrak{I}_{rigflow}/\hbar^2$ $(MeV)^{-1}$ | $2\mathfrak{I}_{rigbody}/\hbar^2$ $(MeV)^{-1}$ | **Predicted** $eQ_o$ **/*B*(*E*2)** $e\,fm^2 / e^2\,fm^4$ | **Measured** $eQ_o$ **/*B*(*E*2)** $e\,fm^2 / e^2\,fm^4$ |
|---|---|---|---|---|---|---|---|
| ${}^{8}_{4}Be$ | $2^+$ | 3.6/2.9 | 1.7/2.0 | 3.3 | 5.3 | 25/62 | 19 / $34 \pm 10.5$ |
| | $4^+$ $J_c = 4^+$ | 10.5/11.4* | 2.0/1.7 | 3.6 | 5.7 | 40/81 (19/- at *J*=0) | Qo is for *J*=0 (*HF-BCS* pre-diction for $2^+$ [52]) |
| ${}^{12}_{6}C$ | $2^+$ | 4.7/4.4 | 1.3/1.4 | 2.5 | 6.0 | -11/12 | -21$\pm$10.5 /11-99 Qo is for *J*=0 [40] |
| | $4^+$ $J_c = 4^+$ | 14.0/14.1* | 1.5/1.5 | 2.7 | 6.4 | 7/3 (-17.7/- at *J*=0) | / 8.5 for $2^+$ [42,43] |
| ${}^{20}_{10}Ne$ | $2^+$ | 1.4/1.6 | 4.4/3.7 | 7.1 | 28.6 | 54/290 | $70 \pm 17.5\vert_{J=0}$ /274-762 [40], $57 \pm 8$ [43], $480 \pm 8$ [53], |
| | $4^+$ | 4.4/4.3 | 4.6/5.4 | 7.3 | 29.2 | 67/233 | / $71 \pm 7$ [43] |
| | $6^+$ | 8.8/8.8 | 5.0/4.9 | 7.6 | 30.3 | 92/86 | / $66 \pm 8$ [43] |
| | $8^+$ $J_c = 8^+$ | 13.8/12.0* | 5.0/9.5 | 8.2 | 32.8 | 153/997 (49/- at *J*=0) | / $24 \pm 8$ [43] |
| ${}^{24}_{12}Mg$ | $2^+$ | 1.3/1.4 | 4.6/4.4 | 7.3 | 35.0 | 30/91 | $84 \pm 21\vert_{J=0}$ /395-1097 [40] $119.3 \pm 25$ [43], $425 \pm 29$ [53] |
| | $4^+$ | 4.3/4.1 | 4.8/5.1 | 7.4 | 35.7 | 46/105 | / 95, + 21, -16 [43] |
| | $6^+$ | 8.5/8.1 | 5.2/5.5 | 7.7 | 36.9 | 74/245 | / 140, + 193, - 49 [43] |
| | $8^+$ $J_c = 8^+$ | 13.6/13.2* | 6.0/5.9 | 8.2 | 39.4 | 134/765 (24/- at *J*=0) | / 74, + 148, - 29 [43] |
| ${}^{28}_{14}Si$ | $2^+$ | 1.5/1.8 | 4.0/3.4 | 7.3 | 40.7 | -53/277 | $-38.5 \pm 21\vert_{J=0}$ /30.5-352.2 [40], $72 \pm 9$ [43], $317 \pm 17$ [53] |
| | $4^+$ | 4.1/4.6 | 4.4/4.9 | 7.6 | 42.5 | -7/3 | / $96 \pm 8$ [43] |
| | $6^+$ $J_c = 6^+$ | 8.8/8.5* | 5.4/5.6 | 8.3 | 46.6 | 103/482 (-70/- at *J*=0) | / $106 \pm 55$ [43] |

***** No measured ground-state rotational-band energy level above this energy is reported in the Table of Isotopes and Nuclear Data Sheets.

**Table 2. Predicted/measured excitation energy ( $\Delta E_J$ ), Cut-off *J* ( $J_c$ ), moments of inertia ( $\mathfrak{I}_J$ ), $eQ_o$ , *B*(*E*2)**

| | *J* | $\Delta E_J$ (*MeV*) model/exp | $2\mathfrak{I}_J/\hbar^2$ $(MeV)^{-1}$ model/exp | $2\mathfrak{I}_{rigflow}/\hbar^2$ $(MeV)^{-1}$ | $2\mathfrak{I}_{rigbody}/\hbar^2$ $(MeV)^{-1}$ | Predicted $eQ_o$ /*B*(*E*2) *e fm*$^2$ / w.u.[7] | Measured $eQ_o$ /*B*(*E*2) *e fm*$^2$ / w.u. |
|---|---|---|---|---|---|---|---|
| $^{162}_{66}Dy$ | $2^+$ | 0.08/0.09 | 76/69 | 89 | 2892 | 334/213 | -/ $204 \pm 3$ [54] |
| | $4^+$ | 0.27/0.27 | 76/78 | 89 | 2894 | 353/122 | -/ $289 \pm 12$ [54] |
| | $6^+$ | 0.56/0.55 | 76/78 | 89 | 2897 | 382/126 | -/ $301 \pm 17$ [54] |
| | $8^+$ | 0.95/0.92 | 76/81 | 90 | 2901 | 422/146 | -/ $346 \pm 17$ [54] |
| | $10^+$ | 1.45/1.38 | 77/84 | 90 | 2907 | 475/180 | -/ $350 \pm 23$ [54] |
| | $12^+$ | 2.04/1.90 | 77/88 | 90 | 2914 | 542/229 | -/ $330 \pm 40$ [54] |
| | $14^+$ | 2.74/2.49 | 78/91 | 90 | 2922 | 625/301 | -/ $330 \pm 40$ [54] |
| | $16^+$ | 3.53/3.14 | 79/96 | 91 | 2933 | 727/402 | N/A |
| | $18^+$ | 4.41/3.83 | 79/101 | 91 | 2946 | 843/1550 | N/A |
| | $20^+$ | 5.38/4.58 | 81/105 | 92 | 2963 | 1013/771 | N/A |
| | $22^+$ | 6.43/5.35 | 82/111 | 92 | 2985 | 1228/1128 | N/A |
| | $24^+$ $J_c = 14^+$ | 7.54/6.15 | 85/117 | 93 | 3022 | 1586/1875 (327/- at *J*=0) | N/A |
| $^{168}_{68}Er$ | $2^+$ | 0.08/0.08 | 75/75 | 99 | 3310 | 367/244 w.u.[8] | -/ $213 \pm 4$ [55] |
| | $4^+$ | 0.26/0.26 | 76.0/76.0 | 99 | 3320 | 424/167 w.u. | -/ $319 \pm 9$ [55] |
| | $6^+$ | 0.55/0.55 | 77/77 | 99 | 3336 | 517/220 w.u. | -/ $424 \pm 18$ [55] |
| | $8^+$ | 0.94/0.93 | 78/79 | 100 | 3359 | 653/333 w.u. | -/ $354 \pm 13$ [55] |
| | $10^+$ | 1.41/1.40 | 80/81 | 101 | 3392 | 843/538 w.u. | -/ $308 \pm 13$ [55] |
| | $12^+$ | 1.97/1.94 | 83/84 | 102 | 3438 | 1116/925 w.u. | -/ $345 \pm 18$ [55] |
| | $14^+$ $J_c = 14^+$ | 2.59/2.57* | 87/86 | 105 | 3512 | 1563/1791 w.u. (343/- at *J*=0) | -/ $336 + 20, - 69$ [55] |

***** No measured ground-state rotational-band energy level above this energy is reported in the Table of Isotopes and Nuclear Data Sheets.

[7] In weisskopt unit, 1 w.u. = 52.3 $e^2\ fm^4$ **.**

[8] In weisskopt unit, 1 w.u. = 54.9 $e^2\ fm^4$ **.**